\documentclass[twocolumn,showpacs,nofootinbib,10pt]{revtex4-1}

\usepackage{color,float,amssymb,amsmath,upgreek}
\usepackage{graphicx}
\usepackage[dvipsnames]{xcolor}
\usepackage{graphicx}\usepackage[colorlinks,hyperindex]{hyperref}
\usepackage[OMLmathrm,OMLmathbf]{isomath}

\newcommand{\beq}{\begin{eqnarray}}
\newcommand{\eeq}{\end{eqnarray}}
\newcommand{\be}{\begin{equation}}
\newcommand{\ee}{\end{equation}}
\newcommand{\bea}{\begin{eqnarray}}
\newcommand{\eea}{\end{eqnarray}}

\usepackage{amsmath}
\usepackage{amsfonts}
\usepackage{amssymb}
\usepackage{amsxtra}
\usepackage{graphicx}
\usepackage{caption}
\usepackage{bookmark}
\usepackage{hyperref,color,xcolor}
\hypersetup{hidelinks,backref=true,pagebackref=true,hyperindex=true,colorlinks=true,breaklinks=true,urlcolor= blue}
\hypersetup{%
  colorlinks = true,
  linkcolor  = blue
}

\newcommand{\ba}{\begin{array}}
\newcommand{\ea}{\end{array}}

\begin{document}

\title{Gregory--Laflamme analysis of MGD black strings}

\author{A. Fernandes--Silva}
\email{armando.silva@aluno.ufabc.edu.br}
 \affiliation{
Centro de Ci\^encias Naturais e Humanas, Universidade Federal do ABC -- UFABC, 09210-580, Santo Andr\'e, Brazil.
}

\author{R.~da~Rocha}
\email{roldao.rocha@ufabc.edu.br}
\affiliation{Centro de Matem\'atica, Computa\c c\~ao e Cogni\c c\~ao, Universidade Federal do ABC -- UFABC, 09210-580, Santo Andr\'e, Brazil.}

%
%


\begin{abstract}

The minimal geometric deformation (MGD), associated with the 4D Schwarzschild solution of the Einstein equations, is shown to be a solution of the pure 4D Ricci quadratic gravity theory, whose linear perturbations are then implemented by the Gregory--Laflamme eigentensors of the Lichnerowicz operator. The stability of MGD black strings is hence studied, through the correspondence between their Lichnerowicz eigenmodes and the ones associated with the 4D MGD solutions. Its is  shown that there exists a critical mass driving the MGD black strings stability, above which the MGD 
black string is precluded from any Gregory--Laflamme instability. The general relativistic limit leads the MGD black string to be unstable, as expected, corresponding to the standard Gregory--Laflamme black string instability. 
\end{abstract}



\maketitle

\section{Introduction}

Recent enterprises to establish theories of gravity, in a scope that is beyond the general relativity (GR), have been 
reasonably successful attempts  to describe our universe. The prototypical  Einstein--Hilbert action is well known to govern the GR Einstein equations. Adding terms to such an action, that are  beyond the  scalar curvature,  does not immediately imply that the derived equations of motion are reduced to the  GR Einstein equations,  even at low energy regimes. 
A recently proposed setup, known as the minimal geometric deformation (MGD) \cite{Ovalle:2017fgl,ovalle2007,ovalle2008,covalle3},   has been shown to be a successful framework procedure to generate 
realistic analytical solutions  of the effective Einstein field equations, by deforming GR solutions \cite{ovalle2007,ovalle2008}. The MGD procedure can  provide new compact stellar distributions \cite{covalle3,Casadio:2012pu,Ovalle:2013vna,Ovalle:2013xla}, with solid crusts due to a bulk Weyl fluid 
\cite{darkstars}, that might collapse into black holes. This procedure is usually implemented by considering more intricate forms of the stress-energy tensor, that are beyond the ones describing perfect fluids in GR \cite{Ovalle:2016pwp,Ovalle:2013vna}.  The MGD setup can be established for anisotropic spacetime fluids and other interesting examples already paved in the literature \cite{Ovalle:2017fgl}. 

The key point of the MGD scenario was originally derived on brane-world models,  wherein the 4D universe consists of the brane itself, that is a solution of an effective string action, placed into a 5D AdS bulk \cite{Antoniadis:1998ig,Antoniadis:1998ig1,lisa2}. 
The brane has finite self-energy described as the brane tension $\sigma$,  which can fluctuate. At low energy regimes, the brane tension emulates a chemical potential characterizing the energy expense for, at least hypothetically, creating  any unit brane volume \cite{sinha}. An infinite value of the brane tension dictates a perfectly rigid brane, corresponding to the GR limit. The  brane tension can be approximated by a constant parameter related to the brane self-energy or an effective 4D brane cosmological constant, in an equilibrium brane configuration. However, more realistic scenarios, that further incorporate brane-world cosmology and includes the universe expansion and the reheating era as well, are  described  by variable tension branes \cite{GERGELY2008,PRD,Bazeia:2014tua}. Moreover, the brane tension has, still,  zero mean local  fluctuations, that are position dependent.

The MGD method is essentially based upon how 5D bulk constituents, encrypted in the 5D bulk Weyl tensor describing a Weyl fluid, can induce physical effects on the 4D brane. In fact, the 5D bulk Weyl fluid percolating the brane implements the MGD setup, responsible to deform the radial component  of any static,  spherically symmetric,  metric, incorporating the inverse of the brane tension $\sigma^{-1}$. MGD solutions are led to the 4D Schwarzschild solution at low energy regimes, when $\sigma\to\infty$. 

One of our aims here is to provide an alternative description that does not make necessary the existence of a 5D bulk Weyl fluid, whose effects can be, for our purposes, emulated by a 4D modified Ricci quadratic gravity setup. 
In fact, the Sasaki-Shiromizu-Maeda procedure makes the 5D bulk pure gravity, described by the 5D Einstein equations with cosmological constant,  to be projected onto the 4D brane \cite{maedas}. The Gauss-Codazzi method then splits the 
5D stress-energy tensor  into a 4D energy-momentum tensor,  
that consists of terms describing, respectively, the energy and matter on the brane, Kaluza-Klein signatures from the bulk, and high energy corrections due to the 5D Weyl fluid \cite{maedas,maartens}. The paradigm throughout the present paper relies on 
the perturbing the Schwarzschild-like solutions in 4D Ricci quadratic gravity \cite{lich17}, alternatively dismissing the necessity of a 5D Weyl fluid
to describe the MGD solutions. Hence, this setup provides the study of 
 MGD black holes and the stability of MGD black strings.

 Standard black strings were shown to be unstable to long wavelength perturbations \cite{Gregory:1993vy,Konoplya:2008yy,Konoplya:2011qq}. \textcolor{black}{The first attempt to localise the black string
close to the brane appeared in Refs. \cite{Kanti:2001cj,Kanti:2003uv}
}. Gregory--Laflamme  instabilities in the black string are related to the Rayleigh--Plateau instability, in fluid mechanics \cite{Choptuik:2003qd}. However, we shall show that there is a critical black string mass, above which the MGD black strings can be stable under long wavelength perturbations.  
 Indeed, besides placing the MGD procedure, complementarily, as a legitimate 4D solution of the Ricci quadratic gravity theory, this paper is also devoted to study the stability of the MGD black strings under Gregory--Laflamme perturbations. In a previous work, the MGD black string was studied, showing that the brane tension robustly influences its event horizon \cite{Casadio:2013uma}. The MGD scenario is suitable for the study of black strings, as brane 4D black holes can be produced by gravitational collapse \cite{Chamblin:1999by}. The limitations of the  approach in Ref. \cite{Casadio:2013uma} regard the near-brane approximation of
the black string  \cite{Bazeia:2014tua,daRocha:2013ki}, whose global form for arbitrary distances from the brane into the bulk, cannot be entirely provided.
Although globally limited by the range of validity of a Taylor expansion 
into the bulk, the MGD black string revealed in Ref. \cite{Casadio:2013uma} prominent features, beyond the standard (Schwarzschild) black string \cite{maartens,Seahra:2004fg}. In fact, for some eras of the universe, the MGD black string event horizon  was shown to  have a finite range, since it has a collapsing throat in the bulk that collapses  to a point along the extra dimension  \cite{Casadio:2013uma}.

More precisely, generalized black strings features  have been widely investigated, mainly in realistic brane-worlds with variable tension \cite{Bazeia:2014tua,HoffdaSilva:2009ht}, also in the context of the MGD extensions \cite{Casadio:2013uma,Casadio:2015gea}. However, all these approaches are valid near the brane, for a Taylor expansion of the brane metric onto the bulk is a good approximation for the purposes therein. For particular cases, the perturbative setup  provides very interesting and relevant solutions for bulk regularity \cite{Bazeia:2014tua,HoffdaSilva:2009ht,CoimbraAraujo:2005es}. 
The Plateau--Rayleigh instability was also emulated in the MGD black string in Ref.
\cite{Casadio:2013uma}. Near enough to the brane, the Taylor expansion of the bulk metric -- as a function of the variable brane tension, the 5D bulk Weyl tensor and the 5D cosmological constant as well -- is an excellent approximation that provides  the black string dripping into higher entropy 5D black holes states, after the formation of a black string throat along the extra dimension at finite cosmological times, also providing hints for its final state.  Nonetheless, this  phenomenon can just occur in a region of the bulk that is very near the brane to make the Taylor expansion to have a controlled error. Ref. \cite{Casadio:2013uma} was the first to study the MGD black string, which was shown to have a finite extent along the extra dimension. Due to its intrinsic features, the original MGD procedure originated a MGD black string, whose analysis near the brane was an exact approach for its global behaviour  \cite{Casadio:2013uma,Casadio:2015gea}, owing to a black string throat that is constricted by the 
variable brane tension as the 4D universe expands, reaching a event horizon with zero radius at finite cosmological time. Although valid for specific eras of the 4D universe, a final and general solution for the MGD black string stability problem lacks, still. 

Here  a complementary paradigm for performing  
the Gregory--Laflamme procedure shall be used, to derive the MGD black string stability 
analysis. Besides, the MGD solution is here shown to be a solution in modified gravity. 
The pivotal point to study the MGD black string stability  
is the Gregory--Laflamme procedure itself \cite{Gregory:1993vy,Gregory:2000gf,Wiseman:2002ti}. The method to be used is a correspondence 
between fluctuations about a 5D black string solution, which was shown to be related to black hole instabilities in 4D massive gravity theory and to black hole instabilities in 4D higher-derivative gravity theories \cite{Babichev:2013una,Brito:2013wya,Myung:2013doa}. 
Besides, black strings can be further be placed into the membrane paradigm, where the fluid/gravity correspondence plays a prominent role \cite{Matsuo:2013qoa,daRocha:2017lqj}. 

In order to study the MGD setup as a solution of 4D actions for the gravity, more intricate actions, beyond the  Einstein-Hilbert one, can involve higher order derivatives, being obstructed by the  Ostrogradsky theorem, proving the instability of some higher derivative theories of gravity. The ones that circumvent this theorem can play prominent roles as theories of gravity, due to their renormalizability, as for example the 4D Gauss-Bonnet Lagrangian, whose spacetime integral is a topological invariant. Ref. \cite{lich17} introduced invariants constructed upon quadratic terms involving the Ricci tensor and  scalar curvature. 
 
Negative-eigenvalue eigentensors of the Lichnerowicz operator play a central role
in the analysis both of static 4D non-Schwarzschild black hole solutions 
\cite{Lu:2015cqa,Lu:2015psa} and also in the analysis of time-dependent black-hole instabilities. Gregory--Laflamme instabilities \cite{Gregory:1993vy} regarding fluctuations around MGD black string solutions are going to be studied in the context of black hole instabilities
 in 4D dimensional higher-derivative gravity theory \cite{Myung:2013doa}. 
However, in this work we shall not study the Gregory--Laflamme instabilities via bulk effects, but rather from their black hole effects on the brane using Ricci quadratic gravity theory. Linear perturbations around  black hole solutions of massive gravity theories exhibit an unstable mode analogous to the Gregory--Laflamme instability of Schwarzschild black strings \cite{Babichev:2013una}. 

This paper is organized as follows: in Sect. II, the MGD is briefly reviewed and Sect. III is devoted to the study of the eigenmodes of the Lichnerowicz operator, in a Gregory--Laflamme analysis of 
the MGD black holes, constructed upon perturbing the Schwarzschild solution, using the (inverse of the) brane tension as the parameter driving the perturbation, originating the MGD solutions in the 4D 
Ricci quadratic gravity. The associated Lichnerowicz eigenvalue problem is split into  time-dependent transverse traceless modes, being lead to a 1D radial wave equation with a potential, in tortoise coordinates. The results 
in a variable brane tension scenario are then compared to the GR limit. Sect. IV describe the Gregory--Laflamme analysis of the MGD black strings.
Sect. V is dedicated to the concluding remarks and perspectives. 

\section{Minimal Geometric Deformation (MGD)}

The MGD procedure shall  be briefly reviewed, in the original setup that induces high energy corrections to the GR on a 4D brane, percolated by a bulk Weyl fluid  \cite{Ovalle:2017fgl,ovalle2007,ovalle2008,covalle3}. 
Greek indexes 
run through spacetime (brane) quantities, whereas capital roman indexes denote the  bulk ones. 
The 4D Einstein  equations encode an effective approach, when the  Gauss--Codazzi method is applied to the 5D bulk Einstein equations with cosmological constant, $R_{AB}-\frac12{\cal{R}}g_{AB}=-\Lambda_5 g_{AB}$, where $\Lambda_5$ is the 
5D cosmological constant, the $g_{AB}$ denote the bulk metric components, $R_{AB}$ is the Ricci tensor and ${\cal R}=g^{AB}R_{AB}$. \textcolor{black}{The brane tension shall be denoted by $\sigma$, hereon.}
 The projected equations onto the brane, after the Shiromizu--Sasaki--Maeda procedure \cite{maedas}, can be expressed in the standard form 
\be
G_{\mu\nu}  = \kappa_4^2 \mathcal{T}_{\mu\nu}. \label{A1}
\ee 
where $\kappa_4^2$ is the 4D coupling constant and $G_{\mu\nu}=R_{\mu\nu} - \frac{1}{2}{\cal R} g_{\mu\nu}$ is the Einstein tensor, and the effective stress tensor reads, 
\be
\mathcal{T}_{\mu\nu}= \kappa_4^2T_{\mu\nu}+ \frac{\kappa_4^2}{2\sigma} S_{\mu\nu}- 6\mathcal{E}_{\mu\nu}. \label{A2}
\ee 
The brane tension of fluid membranes can explain the tiny value of the cosmological constant, based upon  its small
brane tension fluctuations \cite{sinha}. 
The
analogy is based upon a correspondence between statistical mechanics and quantum field theory.
The description of a brane as a spacetime  corresponds to the geometric portrait  of spacetime itself as a 4D manifold, whose quantum fluctuations do correspond to the brane thermal fluctuations \cite{GERGELY2008,PRD}.
\textcolor{black}{More precisely, in the membrane paradigm, the brane itself emulates the 4D universe wherein we live, having its self-gravity manifesting as a brane tension. The general-relativistic framework is based on 
 an infinitely rigid brane, where $\sigma\to\infty$. However the phenomenological bound for the brane tension $\sigma\gtrsim  3.184\times 10^6$ MeV$^4$ must be satisfied, as shown in Ref. \cite{Casadio:2016aum} in the context of the MGD setup. The model of a fluid brane is better approached by considering an E\"otv\"os fluid brane, for which the fluid dynamical brane tension depends on the brane temperature \cite{gly1,gly2,bulk1,bulk2,european}. In fact, the universe has cooled down as it expanded, and the gradient of temperature is related by the time elapsing, also agreeing with the cosmological  standard model. 
}

The first component $T_{\mu\nu}$ in Eq. (\ref{A1}) denotes the brane matter stress tensor accounting a perfect fluid, $
T_{\mu\nu}= (\varrho +p)u_\mu u_\nu - p g_{\mu\nu}$, 
with $\varrho$ as the density, $p$ the pressure scalar field, $u_\mu$ is the $4$-velocity and $g_{\mu\nu}$ is the brane metric.
The second term of \eqref{A2}, the tensor 
\be
\!\!\!\!S_{\mu\nu}= T_\rho^{\;\rho}T_{\mu\nu}-3T_{\mu\rho}T^{\rho}_{\;\nu}+\frac12 g_{\mu\nu}[3T_{\rho\sigma}T^{\rho\sigma}-T_{\rho}^{\;\rho}T_{\sigma}^{\;\sigma}],
\ee
controls the local dominant  high-energy regime,  when $\varrho\gg\sigma$. Besides, $\mathcal{E}_{\mu\nu}$ represents the bulk Weyl tensor projection onto the brane, being related to eigenmodes of bulk gravitons \cite{maartens}. For spherically symmetric, static, metrics, it reads 
\be
\mathcal{E}_{\mu\nu}(\sigma^{-1}) =\left[ \mathfrak{U}u_\mu u_\nu + \left(\frac{1}{3} + \mathfrak{P}\right)h_{\mu\nu}\right]\,\sigma^{-1}, \label{A4}
\ee
where $\mathfrak{U}$ is the bulk Weyl scalar, $h_{\mu\nu}$ is the  projection tensor along the fluid velocity, and $\mathfrak{P}$ is the  stress tensor.
Let us consider a static and spherically symmetric matter distribution, \be
ds^2 = -A(r) dt^2 + \frac{1}{B(r)} dr^2 + r^2 d\varOmega_2,\label{A5}
\ee
where $d\varOmega_2$ is the areal element of the 2-sphere. 
The spherical coordinates $x^0=ct, x^1 = r, x^2=\theta, x^3=\phi$ shall be used. 
Solving Eqs. \eqref{A1}, using the metric \eqref{A5} and the tensor energy momentum \eqref{A2}, relates the density to the pressure of a fluid. Besides, the radial and tangential components of the effective pressure of the Weyl fluid read \cite{ovalle2007,ovalle2008,covalle3}
\beq
\mathring{p}_a &=&  p + \sigma^{-1}\Big(\frac{\varrho^2}{2} + p\varrho + 2 \mathfrak{U} +(-3+7^a) \mathfrak{P} \Big), \label{A11}
\eeq
where $a=0,1$.  Effects of the 5D Weyl fluid  \eqref{A11}  produce anisotropy in the projected fluid on the brane, 
\be
\mathring{p}_0 - \mathring{p}_1 + {6 \mathfrak{P}}{\sigma^{-1}}=0,\label{A6}
\ee
which arises from the difference between the components of the pressure of the fluid, proportional to the inverse of the brane tension $\sigma^{-1}$. Hence, the isotropy of the perfect fluid is restored in the GR limit $\sigma^{-1}\to0$, namely $\mathring{p}_r - \mathring{p}_t=0$, as can be forthwith read off Eqs. \eqref{A2} and (\ref{A4}), yielding Eq. \eqref{A2} to read $\mathcal{T}_{\mu\nu}= T_{\mu\nu}$. 
In order to obtain the deformation term in the metric, the temporal  component 
\beq
g_{00}(r) (=g_{tt}(r))=A(r) = 1-\frac{2GM}{c^2r}\label{A00}\eeq is fixed, to deform the radial component $g_{11}(r) (=g_{rr}(r)) = B(r)$, for $r>R$
\be
\!\!\!\!\!B(r)\!=\!\left(1\!-\!\frac{2GM}{c^2 r}\right)\!\!\left[1\!+\! \frac{R\varsigma}{4r}\frac{\frac{3GM}{c^2R}\!-\!2}{\frac{GM}{c^2R}\!-\!1}\left(1\!-\!\frac{3GM}{2c^2 r}\right)^{\!\!-1}  \right], \label{A9}
\ee 
where $M$ is the  stellar distribution Misner-Sharp mass, $R$ is average radius of the distribution and the parameter $\varsigma$ encodes a bulk-induced deformation of the vacuum, at the compact distribution surface $r=R$ \cite{ovalle2007,covalle3}. 
The MGD parameters are observationally constrained by the classical tests of GR \cite{Casadio:2015jva} and  here the particular case where  \begin{eqnarray}
\label{cobsigma}
\varsigma(\sigma, R)
= 
-\frac{b}{R^2}\sigma^{-1}
\ ,
\end{eqnarray} shall be assumed, for $b\sim  1.348$ a constant parameter  \cite{covalle3,darkstars,Casadio:2015jva}. 
The negativeness of $\varsigma$ compels the  MGD star gravitational field to be weakened, as an effect of a finite brane tension, $\sigma$, when compared to the GR regime $\sigma\to\infty$. Since 
$\varsigma \sim \sigma^{-1}$ by Eq. (\ref{cobsigma}), Eq. \eqref{A9} can be written as \beq
B(r) = A(r) + \sigma^{-1} f(r)\eeq for some radial function $f(r)$, as indeed expected by the MGD procedure, since at the low energy regime $\sigma ^{-1} \to 0$, the Schwarzschild solution of the GR is recovered. Eq. \eqref{A9}  shows that the $B(r)$ function admits two coordinate singularities,
\be
R_1 = \frac{2GM}{c^2}, \quad  \quad R_2 =\frac{3GM}{2c^2} - \frac{\varsigma}{4}\frac{\left(\frac{3GM}{c^2R}-2\right)}{\frac{GM}{c^2R}-1}.\label{A13}
\ee
\textcolor{black}{Just $R_1$ is a physical singularity.
The scalar curvature, evaluated at $R_2$, is finite.  Moreover, since the solution is an analytical one up to the order $\mathcal{O}(\sigma^{-2})$, the horizon $r_1$ shall not be a physical singularity when higher order terms in the brane tension are taken into account. Higher order Kretschmann scalars are also finite for $R_1$ and do not vanish, evaluated at $R_1$. }
According to Ref. \cite{ovalle2007}, it yields $R_1> R_2 $, thus making $R_2$ unaccessible for observation, in the case of MGD black holes.  Meanwhile, many applications were implemented in the MGD setup, mainly in stellar systems, also regarding MGD glueball stars and their improved range of observability at the LIGO and eLISA experiments  \cite{daRocha:2017cxu}, studies of acoustic perturbations in gas fluid flows  \cite{daRocha:2017lqj} and their observability by lensing effects \cite{Cavalcanti:2016mbe} as well.

\textcolor{black}{Introducing the parameter  \beq\label{elll}
\ell=\frac{R}{4}\frac{\frac{3GM}{c^2R}\!-\!2}{\frac{GM}{c^2R}\!-\!1},\eeq} the scalar curvature  for the MGD metric (\ref{A9}) reads
\beq
\!\!\!g^{\mu\nu}R^\rho_{\;\,\mu\nu\rho}=&&\;r^{-2} \!\left(c^2 r\!-\!2 G M\right)^{-3}\!\! \left(c^2
   (r\!+\!\varsigma\ell )\!-\!3 G M\right)^{-2}\nonumber\\&&
   \left[-4 G^2 M^2 \left(4 G M-3 c^2 r\right) \left(c^2 r-3 G
   M\right)^2\right.\nonumber\\&&\left.+2 c^4 \varsigma\ell ^2 \left(c^2 r-2 G M\right)^3+c^2 G M
   \varsigma\ell  \left(-11 c^6 r^3\right.\right.\nonumber\\&&\left.\left.\!\!\!+66 c^4 G M r^2\!-\!140 c^2 G^2 M^2 r\!+\!96
   G^3 M^3\right)\right].
   \eeq
  The next section is, in particular, devoted to place the MGD solution as a 4D one, in the Ricci quadratic gravity, besides studying their stability. 

\section{Lichnerowicz modes and stability analysis of MGD Black Holes in Ricci Quadratic Gravity}

The 4D Ricci quadratic gravity is described by the Lagrangian \cite{lich17}, 
\be
{\cal L} = R + \alpha_1\, R^{\mu\nu} R_{\mu\nu}\!+\!\alpha_2\, R^2,
\label{1}
\ee 
yielding its derived equations of motion to read
\bea
&& R_{\mu\nu} \!-\!\frac12 R g_{\mu\nu} \!+\! 2\alpha_2 R\,\left(G_{\mu\nu}\!+\!\frac14 R\, g_{\mu\nu}\right)
\nonumber\\&&\!+
2\alpha_1 \left(R_{\mu\alpha\nu\beta} \!-\!\frac18 g_{\mu\nu}\, R_{\alpha\beta}\right)
   R^{\alpha\beta}\nonumber\\&&
\!+(2\alpha_2\!+\!\alpha_1)\, (g_{\mu\nu}\, \square \!-\!\nabla_\mu\nabla_\nu)\, R
\!+\! \alpha_1 \square G_{\mu\nu} =0\,,\label{2}
\eea
Letting $R_{\mu\nu}=0$ after accomplishing the variation implies that 
 \cite{lich17}
\bea
&&\delta R_{\mu\nu}  +
 \left[(2\alpha_2+\alpha_1)(g_{\mu\nu}\square -\nabla_\mu\nabla_\nu)-\frac12 g_{\mu\nu}\right] \delta R
  \nonumber\\&&+\alpha_1\left[\square\left(\delta R_{\mu\nu}\! -\!\frac12 g_{\mu\nu}\delta R\right)
       \!+\! 2R_{\mu\rho\nu\sigma}\delta R^{\rho\sigma}\right]=0\,,
\label{5}
\eea
whose trace yields $\delta R -2[3\alpha_2 + \alpha_1]\,\square\delta R=0$. 
These equations have black solutions that can be lead, in particular to the Schwarzschild black hole for $\alpha_1=0=\alpha_2$. Integrating in the range between the black hole event 
horizon and the infinity yields  $\delta R=0$, 
whenever one chooses the correct sign of the term $\square\delta R$ that corresponds to $m_0^2>0$  \cite{lich17}, 
\be\label{mzero}
m_0^2 = \frac{1}{2(3\, \alpha_2+\alpha_1)}\,,
\ee
for the for scalar mode \cite{Nelson:2010ig,Lu:2015cqa}.  Replacing 
$\delta R=0$ into Eq. (\ref{5}) yields 
\be
\Big(\Box_{\tiny{\rm L}}  - \frac1{\alpha_1}\Big) \, \delta R_{\mu\nu}=0\,,
    \label{7}
\ee
where
\be
\Box_{\tiny{\rm L}} f_{\mu\nu} \equiv -\square f_{\mu\nu} - 2
 R_{\mu\rho\nu\sigma}\, \delta f^{\rho\sigma}\label{8}
\ee
is the Lichnerowicz operator acting on any tensor field $f_{\alpha\beta}$.  The tensor $\delta R_{\mu\nu}$ was shown in Ref. \cite{lich17} to be a transverse traceless tensor. Eq. (\ref{7}) provides a transverse traceless eigentensor
 $\psi_{\mu\nu}$ of the Lichnerowicz operator,
\be
\Box_{\tiny{\rm L}}\, \psi_{\mu\nu}= \lambda\, \psi_{\mu\nu}, \, \label{9}
\ee
with eigenvalue $\lambda=1/\alpha_1$. It provides a perturbation of the Schwarzschild solution, that here shall be identified with the MGD solution (\ref{A00}, \ref{A9}),  
whenever the perturbation parameter is identified to the MGD parameter in Eq. \eqref{cobsigma}. 
The constraint $\alpha_1<0$ must be imposed for  massive spin-2 modes  that are not tachyonic modes \cite{lich17}.

The Gregory--Laflamme analysis of the  MGD black strings can be implemented by starting with the derivation of the negative eigenmodes of the Lichnerowicz operator. Ref. \cite{lich17} made explicit the GR limit, $\sigma^{-1}\to0$,  when $B=A$ in Eq. (\ref{A5}), letting 
\be
\psi_{00}= A\, \xi\,,\quad \psi_{11}= A^{-1}\, \chi  \,,\quad
\psi_{ij}= r^2\, \gamma_{ij}\, \psi \,,\label{10}
\ee
where $\gamma_{ij}={\footnotesize{\begin{pmatrix}r^2&0\\0&r^2\sin^2\theta\end{pmatrix}}}$ and the involved functions, except the $\gamma_{ij}$, are $r$-dependent one-variable functions. The requirement $g^{\mu\nu}\, \psi_{\mu\nu}=0$,
yields $
-\xi + \chi  + 2\, \psi =0$. Besides, the transversality condition $\nabla^\mu\psi_{\mu\nu}=0$ implies that \cite{lich17} 
\be
\chi ' + \frac{2}{r}\,
(\chi -\psi )+ \frac{A'}{2A}\, (\xi+\chi ) =0\,.\label{12}
\ee
The Lichnerowicz operator action on the $\psi_{\mu\nu}$ modes, 
\be
\Box_{\tiny{\rm L}} \, \psi_{\mu\nu}= -\square\psi_{\mu\nu} -
          2 R_{\mu\rho\nu\sigma}\, \psi^{\rho\sigma},
\ee
and the Schwarzschild solution limit  $\sigma^{-1}\to0$ with $A(r)=B(r)$ in Eq. (\ref{A5}), together with  Eq. (\ref{10}), yield \cite{lich17}
\bea
\Box_{\tiny{\rm L}}\psi_{00} &=& \left[-A^2\frac{d^2}{dr^2} -\left(A A' +\frac{2A^2}{r}\right)\frac{d}{dr}+\frac{{A'}^2}{2}\right]\,\xi \nonumber\\&&+ \left(\frac{{A'}^2}{2} - A A''\right)\, \chi  -
   \frac{2 A A'}{r}\, \psi \,,\\
\Box_{\tiny{\rm L}}\psi_{11} &=& \left[-\frac{d^2}{dr^2} - \left(\frac{A'}{A} + \frac{2}{r}\right)\,\frac{d}{dr}
  + \left(\frac{{A'}^2}{2A^2} + \frac{4}{r^2}\right)\right]\, \chi  \nonumber\\&&
  +\left(\frac{{A'}^2}{2A^2} -\frac{A''}{A}\right)\, \xi
  +\frac{2}{r}\, \left(\frac{A'}{A} -\frac{2}{r}\right)\, \psi \,,\\
\Box_{\tiny{\rm L}} \psi_{ij}&=& \Bigg\{
\left[ -A\frac{d^2}{dr^2} -\left(A'+\frac{2A}{r}\right)\frac{d}{dr} +
   \left(\frac{4A}{r^2} -\frac{2}{r^2}\right)\,\right] \psi   \nonumber\\&&- \frac{A'}{r}\,\xi
  +\left(\frac{A'}{r}-\frac{2A}{r^2}\right)\, \chi \Bigg\}\, r^2\gamma_{ij}.
\eea
Taking into account $\lambda=1/\alpha_1$ in Eq. \eqref{9},
and solving the tracefree and transversality 
conditions for $\xi$ and $\psi $ yields a radial ODE,   
\bea
&&\!\!\!\!\!\!\!\Big(1-\frac{2GM}{c^2r}\Big)\, \chi '' -
  \frac{24 GM^2-22 c^2 GM r+4 c^4 r^2}{3 c^2 GM r^2-c^4 r^3}
  \chi ' \nonumber\\&& - \frac{8 GM}{r^2 (c^2r-3 GM)} \,
  \chi =  \lambda\, \chi \,.\label{14}
\eea
The transverse traceless Lichnerowicz modes associated with the solutions of Eq. (\ref{14})  of the form (\ref{10}) can be obtained, for large distances, beyond the horizon event. In this limit, Eq. (\ref{14}) reads $\chi ''+\frac4r + \lambda \chi =0$, yielding the perturbation function $\chi =a_1\exp(\sqrt{-\lambda}r)+a_2\exp(-\sqrt{-\lambda}r)$, for $a_1, a_2$ constants. It is related to one renormalizable mode, for $\lambda<0$.  Ref. \cite{lich17}  numerically showed that \beq\label{mode1}
\lambda\approxeq -0.767,\eeq in full compliance  to Ref.  \cite{Gross:1982cv}.

A transverse and
traceless Lichnerowicz mode, with eigenvalue
$\lambda<0$ was shown to yield black hole
solutions of type (\ref{A5}), ramificating from the Schwarzschild black hole ones when 
$\alpha_1=\lambda^{-1}$ \cite{lich17}. Indeed,  transverse 
traceless perturbations of the Schwarzschild metric,
$g_{\mu\nu}\mapsto g_{\mu\nu} + h_{\mu\nu}$ yielding $\delta R_{\mu\nu}=
\frac12 \Box_{\tiny{\rm L}} h_{\mu\nu}$. 

The MGD metric (\ref{A5}) can be implemented in the Ricci quadratic gravity setup by a linear perturbation of the standard GR solution (\ref{A5}), with respect to the inverse of the brane tension,  
\bea
A(r)\mapsto A(r)(1 + \varsigma\, \mathring A(r))\,,\label{ansatz1}\\
B(r)\mapsto A(r) (1 + \varsigma\, \mathring B(r))\,,\label{ansatz}
\eea where $\mathring A$ and $\mathring B$ are functions to be determined a posteriori,  
 and terms $\mathcal{O}(\sigma^{-2})$  are disregarded, due to the 
most recent bounds for the brane tension \cite{Casadio:2016aum}. In fact, the underlying paradigm of the MGD procedure is based upon this consideration, as  terms of order $\mathcal{O}(\sigma^{-2})$ are shown to be phenomenologically irrelevant \cite{Casadio:2016aum}. Hence, the solution (\ref{ansatz1}, \ref{ansatz}) is as analytical as the MGD itself. 
 The perturbation (\ref{ansatz1}, \ref{ansatz}) 
emulates the one in Ref. \cite{lich17} with arbitrary parameter,
but here the inverse of the brane tension is such a function that drives the perturbation. 

Eqs. (\ref{ansatz1}, \ref{ansatz}) are shown in what follows to be equivalent to the MGD metric (\ref{A5}, \ref{A00}). In fact, 
Eqs. (\ref{9}) taken together with $\delta R=0$ yield the following equations:
\beq
\!\!\!\!\!\!\!\!\!\!\!\!2r^2 A\mathring A'' \!+\! r \big(A\!+\!3\big)\mathring A' \!+\!
r\big( A \!+\! 1 \big)(\varsigma' \mathring B\!+\!\varsigma\mathring B') \!+\! 4\varsigma\mathring B\!=\! 0\,,\label{sat1}\eeq and 
\beq
&&4r^2\big(rA \mathring A' +  \mathring B\big) +\alpha_1\Big(2r^2A
\big(3A-1\big)(\varsigma'' \mathring B+2\varsigma' \mathring B'+\varsigma\mathring B'')\nonumber\\&&+r \big(3A^2 - 14A +
5\big)(\varsigma' \mathring B+\varsigma\mathring B')
+ r(A-1) \big(5A-7\big)\mathring A' \nonumber\\&&+ 4\big(A + 1\big) \varsigma  \mathring B\Big)=0\label{sat2}\,,
\eeq
Our aim is to show that the MGD solution (\ref{A00}, \ref{A9}) satisfies the ans\"atzen (\ref{ansatz1}, \ref{ansatz}).  

The parameter $\varsigma\propto\sigma^{-1}$ in Eq. (\ref{cobsigma}), regarding the 
brane tension,  is well known to vary and cannot be taken as truly being  a constant. Successful phenomenological models
are led by  E\"otv\"os fluid branes, which takes into account 
the cosmological evolution and considers a time-dependent variable brane tension that avoids any global anisotropy \cite{GERGELY2008,PRD}. However, E\"otv\"os fluid branes are based on FRW cosmologies, being  ab initio isotropic. In order to encode CMB (cosmic microwave background) anisotropies,  the brane tension can be more generally assumed to be $r$-dependent, without loss of generality, dismissing at a first analysis any angular dependence. In this approach, the brane tension  can be made to depend upon the position of the black hole on the brane, impacting the Lichnerowicz modes. In fact, Refs. \cite{Barrow:2001pi,mm1} show the relationship between bulk graviton massive modes  stresses  the brane  shear anisotropy, using CMB to impose bounds on the anisotropy 
originated from these stresses. CMB anisotropies can be, thus, formulated 
in brane-world models \cite{maartens,mm1,mm2}, where bulk effects can determine the CMB anisotropy.  Since inhomogeneity and anisotropy are observationally small, cosmological effects of anisotropy can be encrypted into the variable brane tension, as it describes the brane self-energy \cite{mm2,mm3}. This setup implements the brane with inflation \cite{Kanno:2003xy}.

 Eqs. (\ref{sat1}) and (\ref{sat2}) are satisfied for the MGD metric (\ref{A5}, \ref{A00}), when 
\beq\label{satisf}
\mathring A(r)&=&0,\\\label{satisf1}
\mathring B(r)&=&\frac{R}{4r}\frac{\left(\frac{3GM}{c^2R}-2\right)}{\left(\frac{GM}{c^2R}-1\right)}\left(1\!-\!\frac{3GM}{2c^2 r}\right)^{\!\!-1}.
\eeq and the bound of the brane tension $\sigma\gtrsim  3.184\times 10^6$ MeV$^4$ is satisfied.  It makes  the $\varsigma$-family of solutions (\ref{ansatz1}, \ref{ansatz}) to be 
given by the MGD solutions (\ref{A00}) and (\ref{A9}), respectively.

\textcolor{black}{The classical tests of GR were used in the context of the MGD setup, to show that 
the parameter $\ell$ in Eq. (\ref{elll}) satisfies the bound $\ell \lesssim (2.80 \pm 3.45) \times 10^{-11}$ \cite{Casadio:2015jva}. Besides, the bound of the brane tension $\sigma\gtrsim  3.184\times 10^6$ MeV$^4$ holds \cite{Casadio:2016aum}. 
Ref. \cite{Casadio:2017sze}  proved that these effects are more feasible for mini-black holes, working as the MGD of the Schwarzschild solution. In fact, taking into account Eqs. (\ref{cobsigma}, \ref{elll}), for a  black hole of mass $10^9 M_\odot$, where $M_\odot$ denotes the solar mass $M_\odot \approx 1.989\times  10^{30}$ kg, for $\ell \lesssim (2.80 \pm 3.45) \times 10^{-11}$ and $\sigma\gtrsim  3.184\times 10^6$ MeV$^4$, it implies that the MGD of the radial Schwarzschild metric component reads $ 
B(r)=\left(1-\frac{2GM}{c^2 r}\right)\left[1+\alpha\left(1\!-\!\frac{3GM}{2c^2 r}\right)^{\!\!-1}\right], $ where $\alpha\approx10^{-13}$, for supermassive black holes. That is the reason why we can perform our perturbed solution. The effects of accretion onto the black hole, due to the MGD, might 
be substantial for mini-black holes, whose analysis was studied in Ref. \cite{Casadio:2017sze}.}

The MGD black hole solutions (\ref{A00}, \ref{A9}) can be identified to the new perturbative black hole solutions (\ref{ansatz1}, \ref{ansatz}) of Ricci quadratic gravity, whenever the perturbation parameter in Eqs. (\ref{ansatz1}, \ref{ansatz}) is regarded as the inverse of the brane tension. Since  the brane tension is a fundamental scalar field playing the role of the brane self-energy, it can be seen as a bath emulating either the cosmological constant or a minimum of a scalar potential that permeates the spacetime \cite{Kanno:2003xy}.  
 
The  stability properties of the black hole solutions can be now explored. The Lichnerowicz negative eigenvalue shall play a key role on the analysis of the stability for black hole perturbations. In fact, the Lichnerowicz problem of time-dependent tensor fluctuation modes around the MGD solution can be emulated from the Schwarzschild one, when the field equations are led to the Schr\"odinger 1D problem with an appropriate potential \cite{Zerilli:1970se}. The black hole perturbations are based upon a temporal dependence, which makes the spherically symmetric ansatz for transverse traceless modes to read \cite{lich17}
\bea
\psi_{00}&=& A\, \xi(r)\, e^{-i\omega t}\,,\quad
\psi_{01}= \zeta(r)\, e^{-i\omega t}\,,\label{16m}\\\psi_{11}&=& A^{-1}\, \chi (r)\,
   e^{-i\omega t}\,,\quad \psi_{ij}= r^2\, \psi(r)\, \gamma_{ij}\, e^{-i\omega  t}\,.\label{16}
\eea 
The traceless and transverse conditions yield non-trivial equations that  can be solved for the functions $\xi$, $\psi $ and $\zeta$ in Eqs. (\ref{16m}, \ref{16}).  The
Lichnerowicz eigenvalue equation $
\Box_{\tiny{\rm L}}\, \psi_{\mu\nu}=\lambda \psi_{\mu\nu}$ 
can be  recast into a wave equation with a potential, \`a la Zerilli \cite{Zerilli:1970se}. 
Introducing the variable $\phi(r)$, defined
by 
\beq
\phi(r) = -i\omega^{-1}\, u_1(r)\, \zeta(r) + u_2(r)\, \psi(r)\,,\eeq as a linear combination of $r$-dependent functions $u_1, u_2$ \cite{lich17}, and taking into account the tortoise coordinate $r_\star\in (-\infty,+\infty)$ in the MGD metric (\ref{A00}, \ref{A9}), 
\beq
r_\star = \int_{r_0}^r \frac{d{\rm r}}{\left(1-\frac{2GM}{c^2{\rm r}}\right)\sqrt{1+\left(1-\frac{3GM}{c^2{\rm r}}\right)^{-1}\varsigma\frac{\ell}{{\rm r}}}}\,,\eeq
the function $\phi(r)$ satisfy a Schr\"odinger-like form of the Lichnerowicz eigenvalue equation,
\be
\left(\frac{d^2}{dr_\star^2} +\omega ^2 - V(r_\star)\right)\phi(r_\star)=0\,,\label{18}
\ee
for the Zerilli-type potential 
\begin{widetext}
\beq
V(r)&=&-\frac{1}{2 c^4 r^5 \left(c^2 r-3 G
   M\right)^3 \left(c^2 (r+\varsigma\ell )-3 G M\right)}\nonumber\\&&\times\Big\{c^{12} r^5 \left(2 l^2 \varsigma  ^2+(\lambda +7) \ell r \varsigma 
   +(\lambda +4) r^2\right)-c^{10} G M r^3 \left(\ell ^2 (19 r+8)
   \varsigma  ^2+\ell  r \varsigma   ((11 \lambda +80) r+16)\right.\nonumber\\&&\left.+2 r^2 ((7
   \lambda +30) r+4)\right)+c^8 G^2 M^2 r^2 \left(2 \ell ^2 (29 r+28)
   \varsigma  ^2+\ell  r \varsigma   ((45 \lambda +347) r+160)+2 r^2 (3 (13
   \lambda +60) r+52)\right)\nonumber\\&&-3 c^6 G^3 M^3 r \left(20 \ell ^2 (r+2)
   \varsigma  ^2+\ell  r \varsigma   ((27 \lambda +226) r+192)+8 r^2 (9
   (\lambda +5) r+22)\right)+648 G^6 M^6\nonumber\\&&+9 c^4 G^4 M^4 \left(8 \ell ^2 \varsigma 
   ^2+2 \ell  r \varsigma   ((3 \lambda \!+\!28) r\!+\!48)\!+\!3 r^2 ((11 \lambda
   +60) r\!+\!48)\right)\!-\!54 c^2 G^5 M^5 (8 \ell  \varsigma  \!+\!r (3 (\lambda
   \!+\!6) r\!+\!28))\Big\},\label{zerilli}\eeq\end{widetext}
   The functions $u_1(r),  u_2(r)$ are then determined as
\beq
u_2(r) &=& \frac{r u_1(r)}{\left(1-\frac{2GM}{c^2r}\right)\sqrt{1+\left(1-\frac{3GM}{c^2r}\right)^{-1}\varsigma\frac{\ell}{r}}}\,,\\
u_1(r) &=& \frac{\left(1-\frac{2GM}{c^2r}\right)\sqrt{1+\left(1-\frac{3GM}{c^2r}\right)^{-1}\varsigma\frac{\ell}{r}}\, r^3}{1 -\lambda r^{3}}\,,\label{here}
\eeq in the MGD context. 

To plot the figure below, natural units are adopted where $G=c=1=M$, without loss of generality, following the computational routine in the \texttt{Mathematica} given  in Ref. \cite{Cruciani:1999zc}.
\begin{figure}[H]
\includegraphics[width=17pc]{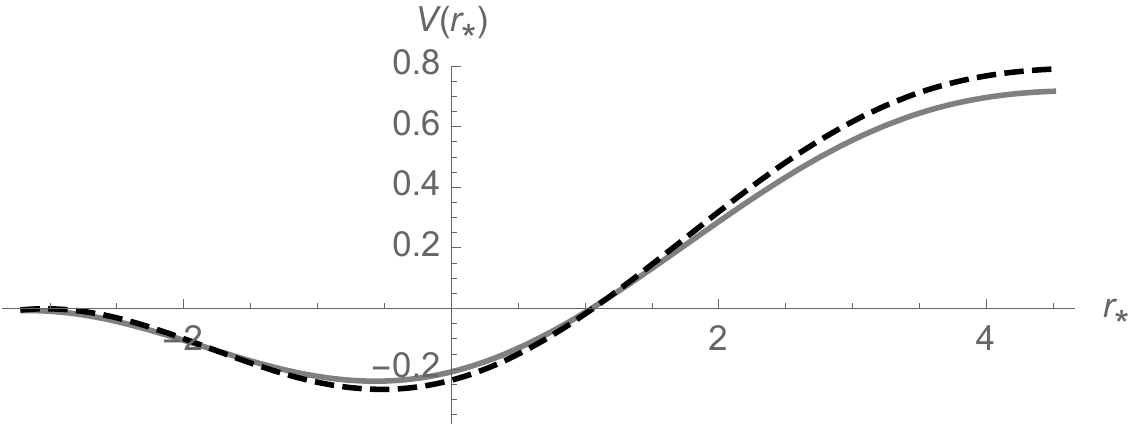}
\caption{\footnotesize{Plot of the Zerilli potential, as a function of the brane tension: the grey line refers to the GR $\sigma\to\infty$ limit, whereas the black dashed line represents the current large value for the brane tension bound $\sigma\sim 3.184\times 10^6$ MeV$^4$.}}
\end{figure}
Following the analysis of the Schwarzschild perturbed black hole in Refs. \cite{lich17,schutz}, a similar qualitative profile is seen in Fig. 1. Although the finite value of the brane tension provides a slightly different plot, with respect to the GR limit $\sigma^{-1}\to0$, the asymptotic limit 
$r_\star\to\pm \infty$ is still similar to the GR limit. 
In fact, when $r_\star\to -\infty$ both plots in Fig. 1 approach to 0, whereas the asymptotic limit when $r_\star\to\infty$ is a constant different of zero.
The phenomenological finite upper bound for the variable brane tension  then alters the exact value of the asymptotic constant for the plot in Fig. 1, when $r_\star\to\infty$. Whilst the qualitative profile of both finite and infinite (Schwarzschild) values of the brane tension are similar, the finite value of the  brane tension 
imposes a different amplitude for the asymptotic outgoing waves $e^{i\omega(t\mp r_\star)}$. 

\section{Gregory--Laflamme analysis of MGD black strings}

The instability of the 4D Schwarzschild solution in Ricci quadratic gravity  is directly related to the 5D black string instability  through the Gregory--Laflamme procedure \cite{Gregory:1993vy}.
Transverse traceless linear perturbations are governed by a negative-eigenvalue mode of the Lichnerowicz operator action on Schwarzschild-type 
backgrounds \cite{lich17,Gross:1982cv}, which was previously analyzed in Eq. (\ref{mode1}). 
Stability properties of the MGD black hole solutions can be derived from the Lichnerowicz negative eigenvalue, that fixes the fringe of the MGD solutions (\ref{A00}, \ref{A9}), seen as a $\varsigma$-family of Schwarzschild black holes (\ref{ansatz1}, \ref{ansatz}), minimally geometric deformed. This is consistent with a variable brane tension, as previously  discussed.

For any possible values of the brane tension $\sigma\gtrsim 3.184\times 10^6$ MeV$^4$, the Gregory--Laflamme procedure yields a family of spherically-symmetric time-dependent
transverse traceless eigentensors with negative eigenvalues $\Lambda$ satisfying \cite{Gregory:1993vy}
\be
\lambda < \Lambda <0\,.\label{23}
\ee
This was accomplished for the Schwarzschild case in the context of the 4D Ricci quadratic gravity. Now,  taking into account the 4D static spherically symmetric metric $ds^2$ in Eq. (\ref{A5}), with the MGD coefficients (\ref{A00}, \ref{A9}), the MGD black string metric reads  $d{\rm s}^2=ds^2 + dy^2$, where one denotes by $y$ the coordinate into the bulk, usually taken as the Gaussian coordinate \cite{maartens,Bazeia:2014tua}. \textcolor{black}{In fact, a black string is, essentially, a higher-dimensional  generalization of a black hole, whose event horizon has  topology  homeomorphic to $S^2 \times S^1$, also having asymptotical topology  $M^{d-1}\times S^1$. A black string is a physical solution in low-energy string theory, described by an extended gravitational object with an event horizon.  \cite{Horowitz:1991cd}.  The prototypical black string metric
is a solution to the Einstein equations in 5D (or even 
higher-dimensional) gravity, being split into a Tangherlini-like black hole and an extra flat
compact dimension. In the context of the brane-world scenarios, the matter trapped onto a brane might undergo 
gravitational collapse, where a black hole with the  extended horizon into the transverse extra direction shall be produced. An observer on the brane shall realize a black hole on the brane, corresponding to a black string in the context of D-dimensional gravity, with metric \begin{equation}
ds^2 = g_{\mu\nu}dx^\mu dx^\nu + dy^2,
\end{equation}
where the $g_{\mu\nu}$ corresponds to the black hole geometry on the brane, and the coordinate $y$ denotes the extra dimension.}

The MGD black string analysis straightforwardly provides the determination of MGD black string instabilities,  in a setup based upon massive gravity theories. In fact, a bulk massive gravity theory has  transverse perturbations\be
\Psi_{\mu\nu} = e^{i\upomega y}\, \psi_{\mu\nu}\,,\label{upo}
\ee
where the $\psi_{\mu\nu}$ are given by Eqs. (\ref{16m}, \ref{16}) and  driven by the Lichnerowicz condition  $
(\Box_{\tiny{\rm L}} + \upomega^2 +m^2)\, \psi_{\mu\nu}=0\,,$ 
where $m$ denotes the mass of the 5D spin-2 bulk gravitons  \cite{Babichev:2013una,Brito:2013wya}. 
It shows the identity for time-dependent fluctuations between the  MGD black string  and the quadratic Ricci gravity, whenever $\upomega^2=-\lambda=-\frac1{\alpha_1}$. The  results assume the form (\ref{upo}),  where the transverse traceless modes are given by Eqs.  (\ref{16m}), (\ref{16}), and \eqref{upo}. 

Taking the metric (\ref{A5}, \ref{A00}, \ref{A9}) of the MGD black hole, with event horizon $\frac{2GM}{c^2}$, the time dependent eigenmodes  of the Lichnerowicz operator exponentially grow, if 
\beq
\frac{\lambda G^2M^2}{4c^4}+\upomega^2+m^2>0.\eeq 
Since $\lambda<0$ by Eq. (\ref{mode1}) \cite{lich17}, 
the criterion for stability reads 
\beq
M^2\gtrsim -\frac{4c^4\lambda}{G^2(\upomega^2+m^2)},\label{cond12}
\eeq
yielding complete stability, for all $\upomega$ to be determined by
$
M^2\gtrsim -\frac{4c^4\lambda}{G^2m^2}. 
$
This condition cannot be satisfied for massless gravity ($m=0$), yielding the standard Gregory--Laflamme instability \cite{Gregory:1993vy}. 
Nevertheless, if MGD black strings in 5D massive gravity theory, then for a MGD black hole event horizon $\frac{2GM}{c^2}$, satisfying (\ref{cond12}), the MGD black string is precluded from a Gregory--Laflamme instability.

\section{conclusions}

We showed that the stability of MGD black strings can be analyzed from the deformation process that is inherent from the MGD construction itself. Different of the Gregory--Laflamme  instability of the classical black string, there is a classical MGD black string mass above which the MGD black string is stable, under long wavelength perturbations. 
The variability of the brane tension is crucial in this point, 
for it to satisfy Eqs. (\ref{sat1}, \ref{sat2}), that cannot hold for constant values of the brane tension. 

The MGD black hole solutions were shown to be solutions of 4D Ricci quadratic gravity.  The Lichnerowicz operator negative eigenvalue rules the stability of the black hole, under perturbations, when its  associated wave equations are identified to the  Schr\"odinger-like 1D problem (\ref{18}), with the potential \eqref{zerilli}. The results are plot in Fig. 1, that shows this potential for the Schwarzschild deformation in Ref. \cite{lich17} and  for the MGD, brane tension-dependent,  solutions, for the maximum value of the brane tension in the phenomenological bound $\sigma\gtrsim  3.184\times 10^6$ MeV$^4$\cite{Casadio:2016aum}. Although the derived profile of both finite and asymptotic values of the brane tension are similar in the two plots in Fig. 1, the finite value of the  brane tension was shown to yield  different amplitudes to the asymptotic outgoing waves, when compared to an infinitely rigid brane.

Eq. (\ref{A13}) determines the event horizons of the MGD standard black hole. As 
the outer event horizon equals the Schwarzschild one, the study of the stability of the MGD black holes cannot be emulated from the Schwarzschild procedure \cite{lich17}, since it yields the same result. This intricate question may be solved and can be implemented through the MGD extensions  \cite{Casadio:2015gea}, that is the most straightforward candidate, beyond the standard MGD procedure itself, to play the role of a different event horizon in the ans\"atzen analogous to Eqs. (\ref{ansatz1}, \ref{ansatz}), for the extended MGD setup. However, this approach has been already revealed to involve hard computational tasks that are being still managed, with relative success \cite{prep1}.
Hence, a more speculative question regards the Bardeen--Cooper--Schrieffer (BCS) and its Cooper instability. In fact, Ref. \cite{hart11} points to strange metals in condensed matter, driven by their duality to 4D Reissner--Nordstr\"om black holes.  Holographic superconductors were shown to be dual to the Reissner--Nordstr\"om black hole, whose   hair is implemented by a Higgs field that composes an atmosphere girdling the black hole, that has an instability shown to be dual to the BCS instability. We then want to answer whether the extended MGD instability can be dual to any condensed matter system.  
  
The next steps also naturally consist of analyzing dilatonic brane-world models
and the deformation of their black hole solutions, where the variable brane tension is 
a scalar field depending upon bulk scalar fields. In this scenario, the AdS/CFT was shown in Ref. \cite{Kanno:2003xy} to be placed into brane-world models, where the effective stress tensor has some terms beyond Eq. (\ref{A2}) that involve the bulk dilaton scalar field and 
its derivatives. This extended   tensor can be, hence, interpreted as the stress-energy  tensor of the CFT matter, whereas the brane tension can be derived, via the Israel-Darmois junction conditions, as the Lie derivative of the bulk scalar field.

\medbreak\bigskip
\paragraph*{{\bf Acknowledgments}. }
The research of AFS is supported by CAPES (Grant 1675676). 
RdR~is grateful to CNPq (Grant No. 303293/2015-2), 
 to FAPESP (Grant No.~2017/18897-8) and to INFN, for partial financial support. Also to Prof. A. Herrera-Aguilar for useful discussions. 

\end{document}